\documentclass[journal]{IEEEtran}
\usepackage{times,amsmath,epsfig,latexsym,amssymb,psfrag,graphicx}
\usepackage{paralist}
\usepackage{booktabs}
\usepackage{color}
\usepackage{setspace}
\usepackage{epstopdf}
\usepackage{subcaption}
\usepackage{mathbbol}
\usepackage[ruled,vlined]{algorithm2e}
\usepackage{citesort}
\newcommand{\qed}{\nobreak \ifvmode \relax \else
\ifdim\lastskip<1.5em \hskip-\lastskip
\hskip1.5em plus0em minus0.5em \fi \nobreak
\vrule height0.75em width0.5em depth0.25em\fi}
\begin{document}

\newcommand{\CS}[1]{\textcolor{magenta}{#1}} 
\newcommand{ \PP}[1]{\textcolor{black}{#1}} 
\newcommand{\AZ}[1]{\textcolor{green}{#1}} 

\hyphenation{auto-regressive superior}

\title{Energy and Resource Efficiency by User Traffic Prediction and Classification in Cellular Networks}
\author{Amin Azari$^*$, Fateme Salehi$^{*\ddag}$, Panagiotis Papapetrou$^\dag$, Cicek Cavdar$^*$\\
$^*$KTH Royal Institute of Technology, Sweden;
$^\dag$Stockholm University, Sweden;
$^\ddag$University of Birjand, Iran \\
Email: $^*$\{aazari, fatemes, cavdar\}@kth.se, $^\dag$panagiotis@dsv.su.se}
\maketitle
 
\maketitle           
\begin{abstract}
There is a lack of research on the analysis of per-user traffic in cellular networks, for deriving and following traffic-aware network management. \textcolor{black}{In fact, the legacy design approach, in which resource provisioning and operation control are performed based on the cell-aggregated traffic scenarios, are not so energy- and cost-efficient and need to be substituted with user-centric predictive analysis of mobile network traffic and proactive network resource management.} Here, we shed light on this problem by designing traffic prediction tools that utilize standard machine learning (ML) tools, including long short-term memory (LSTM) and autoregressive integrated moving average (ARIMA) on top of per-user data.  We present an expansive empirical evaluation of the designed solutions over a real network traffic dataset. Within this analysis, the impact of different parameters, such as the time granularity, the length of future predictions, and feature selection are investigated. As a potential application of these solutions, we present an ML-powered Discontinuous reception (DRX) scheme for energy saving.  Towards this end, we leverage the derived ML models for dynamic DRX parameter adaptation to user traffic. The performance evaluation results demonstrate the superiority  of LSTM over ARIMA in general, especially when the length of the training time series is high enough, and it is augmented by a \textit{wisely}-selected set of features. Furthermore, the results show that adaptation of DRX parameters by online prediction of future traffic provides much more energy-saving at low latency cost in comparison with the legacy cell-wide DRX parameter adaptation.

\end{abstract}

\section{Introduction}\label{sec:introduction}

\IEEEPARstart{P}{roviding} a diverse set of cellular services in an energy- and cost-efficient way is a major driver for the  fifth-generation (5G) wireless networks and beyond (6G) \cite{Sad6g}. To this end, the conventional methods, in which resource provisioning and operation control are performed based on the peak traffic scenarios, are substituted with predictive analytics of mobile network traffic and proactive network resource management \cite{RaUrllc,Db6g,Sad6g}. Indeed, with limited time-frequency radio resources, precise prediction of user traffic arrival in cellular networks can play a key role in improving resource utilization \cite{RaUrllc}. Consequently, in recent years, there has been a growing interest in leveraging machine learning techniques for analyzing the aggregated traffic served in a service area for optimizing the network operation \cite{RnnScale,DrlBsSleep,UavML}. Some exemplary use-cases of traffic prediction in state-of-the-art research works are presented in the sequel. Scaling of fronthaul and backhaul resources by leveraging neural networks for traffic estimation has been investigated in \cite{RnnScale}. Cellular traffic analysis for detecting anomalies in the performance and providing on-demand resources to compensate for such anomalies have been investigated in \cite{UavML}.  Saving energy for base stations (BSs) through prediction of light-traffic periods and sleeping them in the respective periods has been investigated in \cite{DrlBsSleep}. Moreover, light-weight reinforcement learning for computing statistics on interfering packet arrival over different wireless channels has been recently explored in \cite{SelfMl}.

Study of the prior art indicates that while analysis of the aggregated traffic on the network side is an established realm, there is a lack of research on the (i) analysis and understanding at the user level, i.e., analysis of per-user traffic, (ii) prediction and classification of such traffic, and (iii) leveraging such analysis in communications management, especially for energy saving of user devices. Based on the current standardization, DRX (Discontinuous Reception) is the major approach for saving energy of devices in different cellular use-case \cite{moradi2021improving}. DRX implies an inherent tradeoff between reachability and energy saving of devices \cite{ze}, i.e., a longer DRX cycle saves more energy at the cost of non-reachability of the device for a longer period, which on the other hand means longer downlink latency\footnote{In-depth description of DRX and prior arts on DRX configuration will be presented in Section II.B, II.C, and III.B.}. 

In this work, we aim to investigate how analysis of per-user traffic can enable configuration of DRX parameters in a dynamic way to balance the tradeoff between delay and energy saving. To be more specific, given the history of traffic (the IP layer), we are interested in an investigation of (a) how accurately can we estimate (a.1) the intensity of traffic in the next time intervals, (a.2) the application which is generating the traffic?; and (b) how can we adapt DRX parameters of the user device to its traffic? The traffic prediction and classification problems are normally treated as time-series forecasting problems, in which for example, the number of packet arrivals in in a time window is predicted. While the prior art on time-series prediction is mature \cite{TiSerNn,Hyb}, investigation of cellular traffic and communications parameters configuration based on such prediction are challenging tasks \cite{CellTrChar}. Some of the challenges in traffic-aware communications parameter configurations are as follows. First, the traffic per device originates from different applications, e.g., surfing, video and audio calling, video streaming, gaming, etc. Each of these applications could be mixed with another, and could have different modes, making the time series seasonal and mode switching. Second, each application can generate data at least in two modes, in active use and in the background, e.g., for update and synchronization purposes. Third, each user could be in different modes at different hours, days, and months, e.g., the traffic behavior during weekdays differs significantly from the one over weekends. Forth, the features of the traffic, e.g., the inter-arrival time of packets, vary remarkably in traffic-generating applications and activity modes. 

Our preliminary results on generation, labeling, and analysis of cellular traffic captured from a real user have been presented in \cite{azari2019user,azari2019cellular}. Especially, the work in \cite{azari2019user} investigates how autoregressive integrated moving average (ARIMA), as a popular linear predictor,  and long short-term memory (LSTM), as a deep neural network-powered predictor, could be used in the prediction of burst in the traffic from a user device. The work in \cite{azari2019cellular} investigates how the prediction of traffic bursts from a user could be used in radio resource scheduling for reducing latency in communications. In this work, we extend the prior art, and investigate leveraging different ML approaches for saving energy at user devices by configuring the DRX parameters.

\subsection{Paper Contributions and Structure}
\textcolor{black}{While saving energy for end-devices is a mature research area, and many research works have targeted optimization of communications parameters based on the traffic trend, they have a major limitation: dependency to the system parameters, i.e., they can not adapt themselves to the changes in the system, and hence, human intervention is required. On the other hand, artificial intelligence (AI) is leading to an automation revolution in 5G, beyond 5G, and 6G cellular networks.}

\textcolor{black}{Automation of communication parameter configuration has attracted profound attention in recent years \cite{6GArch_automation,netAutomation}. However, in most of recent works, AI has been used for cell-wide parameter configuration, i.e. the aggregated traffic of the cell is predicted, and based on which, cell-wide parameters are tuned \cite{RnnScale,DrlBsSleep,UavML}. In this work, we have done a sanity check to investigate the feasibility of per-user traffic prediction (it is clear that the per-user traffic is much more dynamic and time-varying than the aggregated traffic). Based on the achieved results, we expect such prediction would be of interest in upcoming B5G and 6G applications:
\begin{itemize}
    \item Scheduling of coexisting services
    \item Orchestration of network and radio resources
    \item Configuration of network's and users' communications parameters
    \item Proactive planning of time-frequency resources for URLLC traffic
    \item Battery lifetime-aware serving of IoT communications
    \item etc.
\end{itemize}}

\textcolor{black}{Among these applications, we have selected the third usecase, i.e., configuration of communications parameters, and have shown that indeed, such prediction of traffic could result in considerable saving in energy of devices.
Then, all in all, we see that this work opens up doors to per-user traffic prediction and classification, and its application in configuring communications parameters of devices.}

First, the present work includes an evaluation of features and methods that could be used in the prediction and classification of the traffic, as well as a comprehensive comparative analysis of the tools.  The objective of this comparative study is (i) to investigate how a deep machine learning model compares with a linear statistical predictor model in terms of short-term and long-term predictive performance, and (ii) how additional engineered features, such as the ratio of uplink to downlink packets and protocol used in packet transfer, can improve the predictive performance of the neural network. Second, we investigate the impact of different design parameters and choices, including the length of training data, length of future prediction, the feature set used in machine learning, and traffic intensity, on the performance are investigated. Third, we investigate how a DRX parameter configuration solution could be built on top of this traffic prediction and classification approaches. The major contributions of this work include:
\begin{itemize}
	\item
	Present a per-user DRX parameter configuration scheme leveraging decision trees. 
	Investigate how this adaptation balances the tradeoff between delay and energy consumption.
	\item
	Present a per-user traffic prediction solution, empowered by statistical learning (AR, MA, and ARIMA) and deep learning (LSTM) modules, to be fed to the DRX configuration module. Engineer respective features needed for each module.    
	\item 
	Present a traffic classification solution, empowered by ensemble learning (RAF) and deep learning (LSTM) modules, aiming at specifying the application at the user-side which is generating the traffic, to be fed to the DRX configuration module. Embed the traffic classification module in the overall traffic prediction solution.
	\item 
	Present software and method for making a real labeled  user-side traffic dataset for carrying out traffic prediction/classification. Perform traffic  analysis on the generated dataset, including: (i) performance comparison of  deep-learning against linear statistical-learning in traffic prediction; (ii) performance analysis of adding extra  features to the deep learning predictors; (iii) analysis of performance impacts of design parameters, e.g. the \textit{length} of previous \textit{observations} and future \textit{prediction} on the prediction performance.
	\item
	Identify operation conditions, e.g. the length of training data and the length of future prediction, for which, statistical learning (e.g. ARIMA) outperforms deep learning (e.g. LSTM).
	
\end{itemize}

The remainder of this paper is organized as follows. The next section reviews the related works and the research gap. Section \ref{sec:problem} presents the problem description of traffic prediction and DRX optimization. Section \ref{sec:methods} describes the set of ML methods used throughout the work. Section \ref{sec:experiments} presents the proposed solutions and experimental evaluation results for different solutions as wells as the prediction-powered DRX adaptation. Section \ref{drxa} contains the data-driven DRX adaptation solution, performance evaluation, and discussion on the simulation results. Finally, the concluding remarks and future direction of research are given in Section \ref{sec:conclusions}.

\section{Related Work and Research Gap}\label{sec:related}
In this section, we present the state-of-the-art research works on cellular traffic prediction and classification, and shed light on the research gaps which motivate our research.

\subsection{Cellular Traffic Prediction}
Understanding  dynamics of cellular traffic and prediction of future demands are, on the one hand, crucial requirements for improving resource efficiency \cite{RaUrllc}, and on the other hand, are complex problems due to the diverse set of applications that are behind the traffic. Dealing with network traffic prediction as a time series prediction, one may categorize the state-of-the-art proposed schemes into three categories: statistical learning \cite{bookrima,Seq2Seq}, machine learning \cite{LstmTrafficRaw,SpTmBigDL}, and hybrid schemes \cite{HybArimaLstm}. ARIMA and LSTM, as the flagships of statistical and machine learning approaches for forecasting time series, have been compared in a variety of problems, from economics \cite{ArimaLstm,Seq2Seq,PersonDemand} to network engineering  \cite{NeuTM}. A comprehensive survey on cellular traffic prediction schemes could be found in \cite{DlForecast,EntropyCellular}. A deep learning-powered approach for prediction of overall network demand in each region of cities has been proposed in \cite{NetDemand}. In \cite{SpTmTraffic,SpTmBigDL}, the spatial and temporal correlations of the cellular traffic in different time periods and neighboring cells, respectively, have been explored using neural networks in order to improve the accuracy of traffic prediction.  In \cite{ModelLstmDnn}, convolutional and recurrent neural networks have been combined in order to further capture dynamics of time series, and enhance the prediction performance. In \cite{NeuTM,LstmTrafficRaw}, preliminary results on network traffic prediction using LSTM have been presented, where the set of features used in the experiment and other technical details are missing. Reviewing the state-of-the-art, one observes there is a lack of research of leveraging advanced learning tools for cellular traffic prediction, selection of adequate features, especially when it comes to each user with specific set applications and behaviors.

\subsection{Cellular Traffic Classification}
Traffic classification has been a hot topic in computer/communication networks for more than two decades due to its vastly diverse applications in resource provisioning, billing, and service prioritization, and security and anomaly detection \cite{deepcl,cls2006}.  While different statistical  and machine learning tools have been used till now for traffic classification, e.g. refer to  \cite{lopez} and references herein, most of these works are dependent upon features that are either not available in encrypted traffic, or cannot be extracted in real-time, e.g. port number and payload data \cite{lopez,deepcl}. In \cite{RnnClass}, classification of traffic using convolutional neural network using 1400 packet-based features as well as network flow features has been investigated for classification of encrypted traffic, which is too complex for a cellular network to be used for each user. Reviewing the state-of-the-art reveals that there is a need for investigation of low-complex scalable cellular traffic classification schemes (i) without looking into the packets, due to encryption and latency, (ii) without analyzing the inter-packet arrival for all packets, due to latency and complexity, and (iii) with as few numbers of features as possible. This research gap is addressed in this work.

\subsection{DRX Parameters Configuration} 
DRX provides energy saving at user device by introducing short and long sleep periods during device activity. Using DRX,  user equipment (UE) monitors downlink control channel less frequently, which on the other hand, increased access delay when there are new packets at the BS waiting to be delivered to the UE. Due to the fact of its crucial impact on energy saving, DRX has been embedded in 3GPP LTE standardizations several years ago, and  the literature on DRX modeling and optimizations is mature. Generally speaking, one may categorize the existing literature of DRX into two categories: (a) model-based approaches, and (b) data-driven approach. The majority of prior arts on DRX belong to the former group, in which, (i) traffic arrival to the user is fitted to a probability distribution function (PDF), e.g. exponential distribution, and then (ii), DRX operation is modeled by a Markov process, and finally, optimal DRX parameters are derived based on the parameters of the assumed PDF. The interested reader may refer to \cite{p1,p2,p3,p4}, in which model-based methodology has been used for performance evaluation and optimization of DRX in cellular networks. For example in \cite{p1}, flexibility of DRX in LTE has been investigated and two new DRX modes have been presented. In \cite{p2}, DRX adaptation for autoregressive traffic, also known as self-similar traffic, has been investigated, and by assuming complex analytical models for the traffic, optimized DRX parameters have been found. One main shortcoming of model-based DRX optimization consists in its need for refitting the traffic arrival model to each new stream of data. Due to the large set of installed applications on today's smartphones, and diver set of services like VoIP and video streaming, the traffic arrival model per user can vary in order of minutes, and even seconds. This bottleneck limits application of model-based DRX approaches and has motivated researchers to go beyond model assumptions for traffic arrival.

\subsection{Data-driven DRX Parameters Configuration} 
In recent years, data-driven DRX optimization approaches have gained profound interest in literature. In these schemes, leveraging tools from machine learning, online learning of DRX parameters from the arrival traffic is carried out. To speed up the learning procedure and reduce the complexity of online learning, usually offline learning of traffic prediction models  is done a priory, based on which, the adaptation of the model to  traffic instance is done in an online manner. For example in \cite{p5}, online learning of DRX parameters for IoT devices has been presented. In \cite{p6}, learning from the per-user traffic record is used for the online selection of the best-matched set of DRX parameters. In \cite{p7}, an LSTM-powered deep learning scheme is applied to the user traffic record, which results in an improved prediction of traffic in comparison with the state-of-the-art. This prediction is subsequently employed in the adaptation of one DRX parameter to the traffic. In \cite{wu2021adaptive}, the arrival time of the next packet for IoT devices is predicted at the edge leveraging an ML solution. This prediction is subsequently employed in configuration of DRX parameters for IoT devices. In \cite{ruiz2021drx}, human voice communications are investigated, and a Gaussian process regression algorithm is presented to recognize patterns of silence in the communications. This information is then employed in configuring on/off periods of radio interface for energy saving. In \cite{sundararaju2020novel}, configuration of DRX parameters for devices connected by multiple sim cards has been explored in which, the devices carry out a traffic analysis over its multiple sim cards, and fed the insights to the base station for configuration of the DRX parameters. Flexible DRX configuration for energy/delay performance enhancement in emerging 5G use cases could be found in \cite{moradi2021improving}.

From the above studies, one observes that there is a lack of comprehensive research on cellular traffic prediction and classification (of the application behind the traffic), especially when it comes to the real traffic comprising different traffic types (applications), using different statistical and machine learning tools. Furthermore, utilization of ML-powered prediction and classification approaches in DRX parameter optimization of users is also a less explored area, which is envisioned to be of crucial importance for performance enhancement in emerging 5G/6G use cases \cite{moradi2021improving}.

\section{Problem Description} \label{sec:problem} 
\subsection{Traffic Prediction and Classification} \label{sec:traffic}
In this section, we first introduce the system model,  and formulate the research problem addressed in the paper. Then, we present the overall structure of the traffic prediction framework, which is introduced in this work. The system model considered in this work consists of a cellular device, on which a set of applications are running, e.g., User-1 in Fig. \ref{sys}. 
At a given time interval $[t,t+\tau]$ with length $\tau$, each application could be in an \emph{active} or \emph{background (inactive)} mode, based on the user behaviour. Without decoding the packets, we can define a set of features for aggregated cellular traffic in  $[t,t+\tau]$ for a specific user, such as the overall number of uplink/downlink packets and the overall size of uplink/downlink packets. Let vector $\textbf x(t)$ denote the set of features describing the traffic in the interval $[t,t+\tau]$. Furthermore, let $\textbf{X}_m(t)$ denote a matrix containing the latest $m$ feature vectors of traffic for $m\ge 0$. For example, $\textbf{X}_2(t)=[\textbf{x}(t-1),\textbf{x}(t)]$. Further, denote by $\textbf s$ an indicator vector, with elements either 0 or 1. Then, given a matrix $\textbf{X}_m(t)$ and a binary indicator vector $\textbf s$, we define $\textbf{X}^\textbf{s}_m(t)$ the submatrix of $\textbf{X}_m(t)$, such that all respective rows, for which $\textbf{s}$ indicates a zero value, are removed.  For example, let
\[
\textbf{X}_2(t)=
\begin{bmatrix}
   1 & 2 \\
   3 & 4
\end{bmatrix},
\text{ and      }
\textbf{s}=[1,0] \ ,
\]
then, $\textbf{X}^\textbf{s}_2(t)=[1,2]$.  

Now, the research question in Section~\ref{sec:introduction} could be rewritten as:
%\begin{align} 
%&\textrm{Given}\quad \textbf{X}^{\textbf{s}_1}_{m}(t-1), m\ge 0;\nonumber\\ 
%&\textrm{minimize}\quad L\big(\tilde{\textbf X}_{\text{-}n:\text{-}n\text{-}k}^{\textbf{s}_2}(t),{\textbf X}_{\text{-}n:\text{-}n\text{-}k}^{\textbf{s}_2}(t)\big)\label{opp}\\
%&\textrm{subject to:}\quad \{n,k\}\in \mathbb{Z}, \{n,k\}\ge 0,\nonumber
%\end{align}
%where $[n:n+k]$ is the time window in the future for which, we need the prediction, e.g., $\{n=2, k=1\}$ corresponds to the problem of  prediction of $[\textbf{x}(t+2), \textbf{x}(t+3)]$.  $\textbf{s}_1$  indicates the set of features available in the given dataset, and $\textbf{s}_2$ is the indicator of features set to be predicted. Furthermore, $\tilde{\textbf {X}}_{-n}^{\textbf{s}_2}(t)$ is the prediction of $\textbf {X}_{-n}^{\textbf{s}_2}(t)$. Finally, 
% $L(\cdot)$ is the desired error function, e.g., it may compute the mean squared error  between observations and predictions. 
\begin{align} 
&\textrm{Given}\quad \textbf{X}^{\textbf{s}_1}_{m}(t-1), m\ge 0;\nonumber\\ 
&\textrm{minimize}\quad L\left(\tilde{\textbf X}_{-n}^{\textbf{s}_2}(t),{\textbf X}_{-n}^{\textbf{s}_2}(t)\right)\label{opp}\\
&\textrm{subject to:}\quad n\in \mathbb{Z}, n\ge 0,\nonumber
\end{align}
where $[0:n]$ is the time window in the future for which, we need the prediction, e.g., $n=2$ corresponds to the problem of  prediction of $[\textbf{x}(t), \textbf{x}(t+2)]$.  $\textbf{s}_1$  indicates the set of features available in the given dataset, and $\textbf{s}_2$ is the indicator of features set to be predicted. Furthermore, $\tilde{\textbf {X}}_{-n}^{\textbf{s}_2}(t)$ is the prediction of $\textbf {X}_{-n}^{\textbf{s}_2}(t)$. Finally, 
 $L(\cdot)$ is the desired error function, e.g., it may compute the mean squared error  between observations and predictions.  

Recall the challenges described in the previous section on the prediction of cellular traffic, where the major challenge consists in independency of traffic arrival to user behavior and type of the application(s) generating the traffic. 
 Then, as part of the solution to this problem, one may first predict the application(s) in use and behaviour of the user, and then use them as  extra features in the solution.  In order to realize such a framework, it is of crucial importance to first evaluate the traffic predictability and classificablity using only statistics of traffic with granularity $\tau$, and then, to investigate hybrid models for augmenting predictors by online classifications, and finally to investigate  traffic-aware network management design. In the following sections, predictability and classificablity of real cellular traffic  statistics is investigated, and other parts of the proposed framework are left for the future works.

 \begin{figure}[!htb]
 	\centering
 	\includegraphics[width=1\columnwidth]{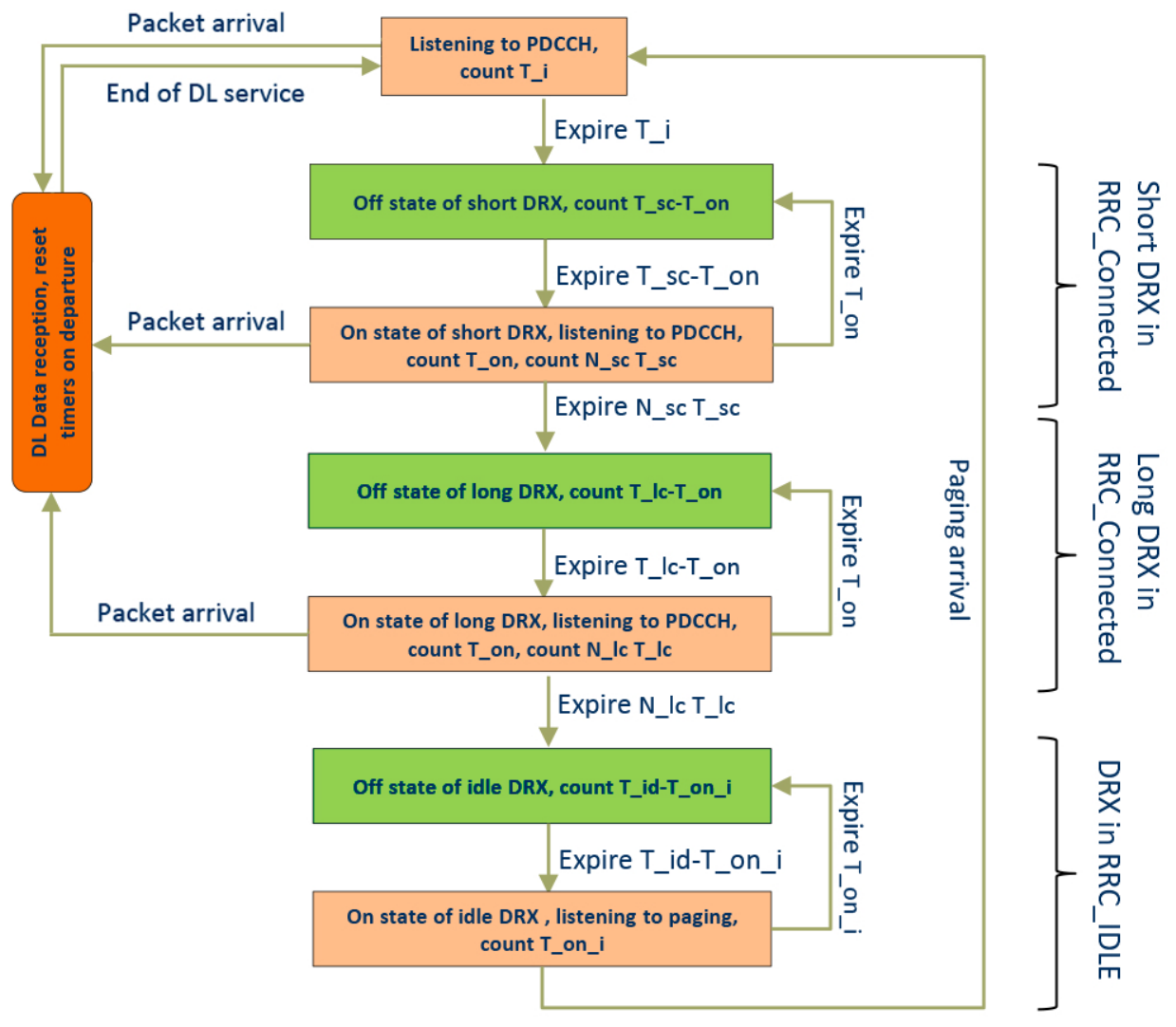}
 	\caption{Different states of activity of UE when DRX is enabled.} 
 	\label{drx}
 \end{figure} 
 
 \subsection{Traffic-adaptive DRX Configuration} \label{sec:DRXprob}
Each user in cellular communications is in 4 states, data transmission, data reception, active, and sleep states. Besides  efforts towards reducing energy consumption in data transmission and reception, reducing energy consumption in the active state has gained profound interest. This is due to the fact that while power consumption in data transmission/reception is higher than the idle state, the spent time in the active state, waiting for potential data, is much higher than the other states. When DRX is used, UE's radio is powered off when there is no data to be transmitted/received. Furthermore, UE wakes up periodically its radio to check if there is any data to be received.

In LTE layer architecture, DRX is controlled by radio resource control (RRC) layer. An LTE device has two modes of operation when a device is switched on, ie, RRC-Idle mode and RRC-Connected mode. In Fig. \ref{drx}, DRX in both RRC-Connected and RRC-Idle modes has been depicted. As we are looking for traffic adaptive DRX optimization, and traffic communications happens in RRC-connected mode, here as per state-of-the-art, we focus on the DRX parameters in the RRC-connected mode. In this mode, there are five DRX parameters, namely, inactivity timer $T_\text{I}$, short DRX cycle $T_\text{sc}$, long DRX cycle $T_\text{lc}$, on duration cycle $T_\text{on}$, and number of short DRX cycles before going to the long DRX cycle $N_\text{sc}$, which must be adapted to the traffic. 

During the short time span $T_\text{I}$, UE monitors physical downlink control channel (PDCCH) for packets in the buffer. If no packet arrives before the expiry of $T_\text{I}$, UE enters to the short DRX state. During short DRX cycle, $T_\text{sc}$, UE sleeps for a short period of time and does not receive/transmit data. This cycle consists of a small ON period, $T_\text{on}$. At the expiry of every short sleep cycle, UE checks for the data during the ON period. The short DRX cycle repeats for $N_\text{sc}$ number of times. After the expiration of this time, i.e., $(N_\text{sc} \times T_\text{sc})$, if no packet arrives, UE switches to the long DRX state. During long DRX cycle, $T_\text{lc}$, UE sleeps for a long period of time. At the expiry of long sleep cycle, UE wakes up and checks for the packets during the ON period $T_\text{on}$. The long DRX cycle is repeated for $N_\text{lc}$ number of times. If no packet arrives before the expiration of this timer, i.e., $(N_\text{lc} \times T_\text{lc})$, UE transits to the idle state.  

Our question in this part could be formulated as follows:
\begin{align}
& \text{Given}\quad \textbf X^{\textbf{s}_1}_m(t-1),\nonumber\\
&\min_{T_\text{I}, T_\text{sc}, T_\text{lc}, T_\text{on}, N_\text{sc}} \mathcal F(D, E),
\end{align}
where $t$ is the decision time, $D$ is the expected experienced delay by data packets during application of the DRX decision ($T$), and is measured from the time packet is added to the buffer at the BS until successful delivery to the device. Moreover, $E$ represents the expected  energy consumption of the device during application of the DRX decision, and $\mathcal F$ represents a function combining the energy and delay indicators. Here for the sake of simplicity, we continue with a weighted sum of the indicators, i.e. $\mathcal F=\omega D+ (1-\omega) E$.

\begin{figure}[!htb]
        \centering
                \includegraphics[width=1\columnwidth]{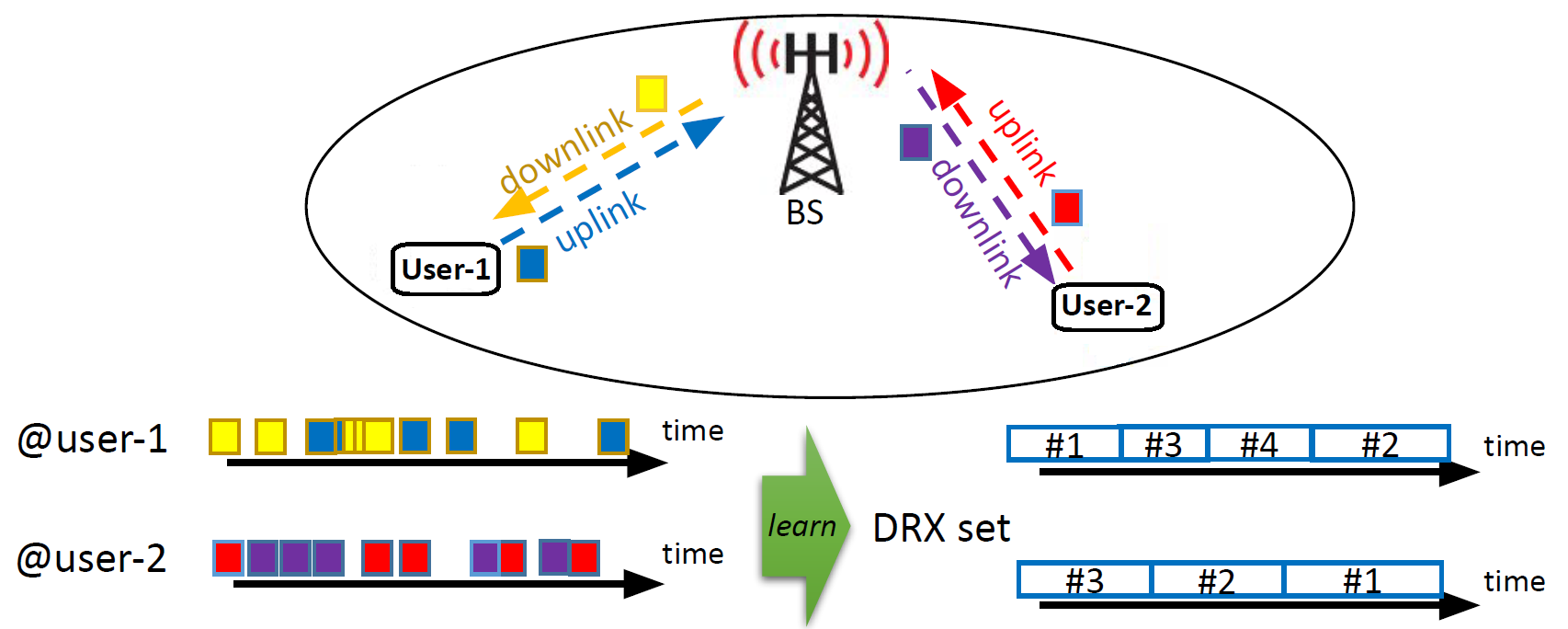}
                \caption{A communication network including base station, users, and data-communication links. DRX sets are dynamically attributed to the users based on the predicted traffic.} 
                \label{sys}
\end{figure}

Fig.~\ref{sys} demonstrates a communication network in which a BS is serving users in the uplink (towards BS) and downlink (towards users), and a sub-intent\footnote{The other sub-intents (with higher/lower priorities) include providing quality of service (QoS), e.g. in terms of latency and reliability.} in communication management is to save energy for devices and radio resource for the network by tuning the DRX parameters. As indicated in this figure, the DRX parameters could be adjusted to some predefined sets if we have a true analysis of per-user traffic in each time interval as well as a true prediction of traffic in the upcoming time intervals. Based on this motivation, the remainder of this paper is dedicated to inspect the feasibility of exploiting the traffic history at the user level and employing it for future traffic prediction via machine learning and statistical learning approaches.

\section{ Traffic Prediction and Classification Solutions} \label{sec:methods}
In this section, we give a short description of the statistical and machine learning solutions used within the proposed prediction framework in Section~\ref{sec:traffic}.  

\subsection{Statistical Learning: AR, MA, and ARIMA} \label{sec:arima}
The first method we consider in our work is Autoregressive integrated moving average (ARIMA), which is essentially a statistical regression model.  The predictions performed by ARIMA are based on considering the lagged values of a given time series, while at the same time accommodating non-stationarity.  ARIMA is one of the most popular linear models in statistical learning for time series forecasting, originating from three models: the autoregressive (AR) model, the moving average (MA) model, and their combination, ARMA \cite{box1976time}. 

More concretely, let $\mathcal{X} = x(1), \ldots, x(n)$ define a uni-variate time series, with $x(i)\in \mathbb{R}$, for each $i \in [1,n]$. A $p$-order AR model, $\text{AR}(p)$, is defined as follows: 
\begin{equation}
\tilde x(t) = c + \alpha_1 x(t-1) + \alpha_2 x(t-2) + \ldots + \alpha_p x(t-p) + \epsilon(t) \ ,
\end{equation}
where $\tilde x(t)$ is the predicted value of the time series at time $t$, $c$ is a constant, $\alpha_1, \ldots, \alpha_p$ are the parameters of the model and $\epsilon_t$ corresponds to a white noise variable.

In a similar, a $q$-order moving average process, $\text{MA}(q)$, expresses the time series as a linear combination of its current and $q$ previous values. Hence, $\text{MA}(q)$ is defined as:
\begin{equation}
\tilde x(t) = \mu +  \epsilon(t) + \beta_1\epsilon(t-1) + \beta_2\epsilon(t-2) + \ldots + \beta_q\epsilon(t-q) \ ,
\end{equation}
where $\mu$ is the mean of $x$, $\beta_1, \ldots, \beta_q$ are the model parameters and $\epsilon(i)$ corresponds to a white noise random variable.

The combination of an AR and an MA process coupled with their corresponding $p$ and $q$ order parameters, respectively, defines an ARMA process, denoted as $\text{ARMA}(p, q)$, and defined as follows:
\begin{equation}
\tilde x(t) = \text{AR}(p) + \text{MA}(q) \ .
\end{equation}
The original limitation of ARMA is that, by definition, it can only be applied to stationary time series.  Nonetheless, non-stationary time series can be stationarized using the $d^{th}$ differentiation process, where the main objective is to eliminate any trends and seasonality, hence stabilizing the mean of the time series. This process is simply executed by computing pairwise differences between consecutive observations. For example, a first-order differentiation is defined as $x^{(1)}(t) = x(t) - x(t-1)$, and a second-order differentiation is defined as $x^{(2)}(t) = x^{(1)}(t) - x^{(1)}(t-1)$. 

Finally, an ARIMA model, $\text{ARIMA}(p, d, q)$, is defined by three parameters $p,d,q$ \cite{mills1991time}, where $p$ and $q$ correspond to the AR and MA processes, respectively, while $d$ is the number of differentiations  performed to the original time series values, that is $x(t)$ is converted to $x^{(d)}(t) =\nabla^d x(t)$, with $x^{(d)}(t)$ being the time series value at time $t$, with differentiation applied $d$ times. 
Consequently, the full $\text{ARIMA}(p, d, q)$ model is computed as follows:
\begin{align}
 \tilde x^{(d)}(t) &= \alpha_1 x^{(d)}(t-1) + \alpha_2 x^{(d)}(t-2) + \ldots \nonumber\\
 &+ \alpha_p x^{(d)}(t-p) +\epsilon(t) + c+ \beta_1\epsilon(t-1) \nonumber\\
 &+ \beta_2\epsilon(t-2) + \ldots + \beta_q\epsilon(t-q) + \mu \ .
 \label{eq:arima}
\end{align}

\begin{figure}[!htb]
        \centering
                \includegraphics[width=1\columnwidth]{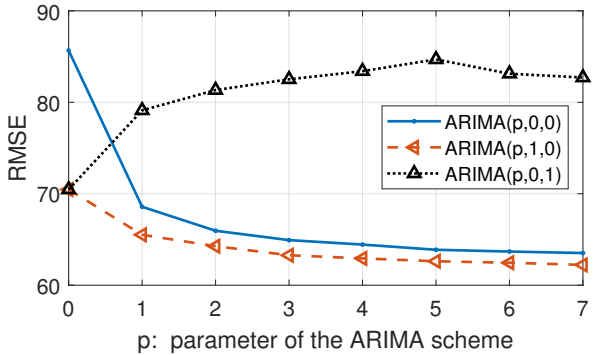}
                \caption{The RMSE performance of ARIMA($p,d,q$) for different $p,d,q$ values. } 
                \label{arimaperf}
\end{figure}

\smallskip
\noindent \textbf{Finding optimized parameters}
In this  study, the ARIMA parameters, including $p$, $d$, and $q$, are optimized by carrying out a grid search over potential values in order to locate the best set of parameters. Fig.~\ref{arimaperf} represents the root mean square error (RMSE) results for different ARIMA($p,d,q$) configurations, when they are applied to the dataset for prediction of the number of future packet arrivals. One observes, among the presented configurations, the optimal performance is achieved by ARIMA(6,1,0) and ARIMA(7,1,0). 

\subsection{Rule-Based Learning: Decision Trees and Random Forests} \label{sec:raf}
A decision tree is a rule-based classifier, where each internal node corresponds to a condition on
a data attribute. The outcome of the condition can be binary, categorical (nominal or ordinal), or real-valued. Depending on the outcome of the condition
the test example follows the corresponding branch, starting from the root node all the way down to a leaf node. Leaf nodes contain a class label, which correspond to the final classification outcome. A path from the root node to a leaf node builds a decision rule. The idea of a single decision tree is extended naturally to random forests (RAF)s and ensemble learning, based on 	the key fact that using an ensemble of many simple weak classifiers can lead to a much stronger classifier, given that each individual weak classifier is slightly stronger than random guessing and independent of all other classifiers. To classify a new object, it is sent to each tree in the forest, and each tree gives a result. The final class label is
determined by majority voting \cite{RAF_ref}. 
More formally, let $h_i$ be a single learner, i.e. in our case a decision tree. Given a data example $x$, the RAF 
determines the final class label as follows using a set of $k$ independent decision trees, as follows: $R(x) = M^{\star} \{h_1 (x), \ldots, h_k(x)\},$ where $M^{\star}$ denotes the majority vote function of the set of individual learners. In this study, RAF is used for traffic classification.

\subsection{Deep Learning: Recurrent Neural Networks} \label{sec:lstm}
The next method we consider in our study is a Recurrent Neural Network (RNN) architecture, using Gated Recurrent Units (GRU) \cite{DBLP:journals/corr/ChoMGBSB14}, which is a generalization of the feed forward network model for dealing with sequential data, with the addition of an ongoing internal state serving as a memory buffer for processing sequences. The key idea is to keep track of arbitrary long dependencies in the input time series.  In this work, we leverage a special class of RNNs, named Long Short-Term Memory (LSTM). 

Let $\{X_1, \ldots, X_n\}$ define a time series of $n$ values, with each $X_i \in \mathbb{R}$. We also let $\tilde{Y}_t$ define the output value produced by the RNN at time point $t$.
The \textcolor{black}{LSTM} architecture \textcolor{black}{we used in this paper is an LSTM with a forget gate $f_t$ and the sigmoid activation function. The architecture} is defined by iterating the following equations:
\begin{eqnarray}
f_t &=& \mathrm{sigm}(W^f X_t + U^f h_{t-1}+b_f) \ ,\\
i_t &=& \mathrm{sigm}(W^i X_t + U^i h_{t-1}+b_i) \ ,\\
o_t &=& \mathrm{sigm}(W^o X_t + U^o h_{t-1}+b_o) \ ,\\
\tilde{c}_t &=& \mathrm{sigm}(W^c X_t + U^c h_{t-1}+b_c) \ ,\\
c_t &=& f_t \circ c_{t-1} + i_t \circ \tilde{c}_t \ , \\
h_t &=& o_t \circ \mathrm{sigm}{c_t}  \ .
\end{eqnarray}
where
\begin{itemize}
    \item \textcolor{black}{$f_t\in (0,1)^h$ is the activation function of the forget gate,}
    \item \textcolor{black}{$i_{t}$ is the input/update gate's activation vector,}
    \item \textcolor{black}{$o_{t}$ is the output gate's activation vector,}
    \item \textcolor{black}{$W^r$, $U^r$, $W^z$, $W^h$, $b\in\mathbb{R}^h$ are the network weight matrices and bias parameters to be learned,}
    \item \textcolor{black}{$h_{t}\in(-1,1)^h$ denotes the hidden state vector,}
    \item \textcolor{black}{$\tilde{c}_t\in(-1,1)^h$ is the cell input activation vector,}
    \item \textcolor{black}{$c_t$ is the cell state vector,}
    \item \textcolor{black}{$h$ and $d$ correspond to the number of hidden units and input features, respectively,}
    \item \textcolor{black}{$\mathrm{sigm}(\cdot)$ denotes the sigmoid function, and}
    \item \textcolor{black}{$\circ$ is the Hadamard product.}
\end{itemize}

Finally, in this paper we perform multi-step RNN-based time series prediction.  That is, given an input time series $\mathcal{X} =\{X_1, \ldots, X_n\}$ running from time point $1$ up until time point $n$ and an expected output series from time point $n+1$ till $m$, i.e., $\mathcal{Y} =\{Y_{n+1}, \ldots, Y_m\}$, the loss function we optimize is the mean squared error, defined over the expected output values within time window $[n+1, m]$ as follows:
\begin{equation}
\mathcal{L} = \frac{1}{m-n}\sum\limits_{t=n+1}^{m} \left(Y_t-\tilde{Y}_t\right)^2.
\end{equation}

\section{Experimental Results of Traffic Prediction and Classification} \label{sec:experiments}
In this section we  investigate the performance of ARIMA and LSTM-powered prediction and classification tools over a real cellular user dataset.

\subsection{Dataset: Generation and Feature Selection } \label{sec:dataset}
For setting up any prediction tool, having access to a large and well-representative dataset is of crucial importance. Reviewing the state-of-the-art, as well as online resources, represents that to the best of authors' knowledge, there are no public  datasets available representing cellular traffic to/from a user. Among several other reasons, privacy is a major reason that results in a lack of availability of cellular traffic records of  users. Then, in order to carry this research out, in this work we generate our own dataset and made part of it available online \cite{GHAA} for future works.

\subsubsection{Generation of the dataset}
 In order to generate the dataset, we leverage a packet capture tool, named "NetGuard: Not-roof firewall", at the user side \cite{ng}. Using this tool, packets could be captured at the Internet protocol (IP) level. One must note that the cellular traffic is encrypted in layer 2, and hence, the payload of captured traffic is neither accessible nor intended for analysis. The latter is due to the fact that for the realization of a low-complexity low-latency traffic prediction/classification tool, we are  interested in achieving the objectives just by looking at the traffic statistics. In this study, two sets of data have been generated: Labeled, and unlabeled data. The former, to be used for traffic predictions, has been generated by logging all outgoing/incoming traffic through the NetGuard app, and during the logging period, user is performing its normal daily activity with the mobile phone. The latter, to be used for classification of the application, is generated in a controlled environment at the user-side in which, we filter internet connectivity   for all applications unless one application, e.g. Skype. Then, the traffic is labeled based on the application running in that time period.
 
\textcolor{black}{There are some limitations on data generation like dependency of it to the applications installed on the devices, e.g., strong presence of Skype for calling, and lack of Twitter on the devices under test. On the other hand, we have done some actions to compensate these shortcomings as much as possible.
First, we have tried to cover 5 different data usage sources of interest, e.g., surfing and video calling, that constitute the main usage of an ordinary person. While we have not included all traffic types, e.g. Twitter, Tiktalk, and etc, but we expect the proposed solutions to be adapt to them.
Second, we have tried to make the length of data collection as long as possible, e.g. from several days to weeks, in different batches of data used in training and testing. One must note that limiting factor in length of our data collection were software problems, lack of space on a device, and etc. For example, when all logs of traffic over an android device is captured by wireshark, after some days of calling and video streaming, Gigs of data are captured on the device, making its hard full of data.}

\textcolor{black}{Based on the fact that our proposed solutions could adapt to different types of traffic in the training phase, we expect that in practice, when a new user with its special traffic characteristics come into the network, the proposed solutions can adapt to it, and work properly with it, as we have analyzed over our own datasets. }

\textcolor{black}{We have used labeling for classification of type of application which is generating the data, e.g. distinguishing surfing from video calling. Towards this end, data capturing is as before, e.g. by a traffic logging software  on the android device, but this time, we write down the time of usage of each application. Then, we import this labeling to wireshark, and shape our dataset.}
  
\subsubsection{Available features in the dataset}
We focus in our study on seven packet features: i) time of packet arrival/departure, ii) packet length, iii) whether the packet is uplink or downlink, iv) the source IP address, v) the destination IP address, vi) the communication protocol, e.g., UDP, and vii) the encrypted payload, where only the first three features are derived without looking into the header of packets. Recall from Section~\ref{sec:traffic}, in which the time has been quantized into intervals of length $\tau$, i.e., the traffic intensity for intervals of length $\tau$ is estimated. It is straightforward to infer that $\tau$ tunes a tradeoff between complexity and reliability of the prediction. In other words, when $\tau$ tends to zero seconds, e.g. $\tau=1$ msec, one can predict traffic arrival for the next $\tau$ interval with high reliability at the cost of extra effort for keeping track of data with such a fine granularity.  On the other hand, when $\tau$ tends to seconds and minutes, the complexity and memory needed for prediction  decrease, which on the other hand also results in less accurate prediction results for the next intervals. In this work, most experiments are carried out for $\tau=10$ sec, which is a practical value. Table~\ref{feat} represents the set of designed features for each time interval in rows, and the subsets of features used in different feature sets (FSs) in different experiments.

\begin{table}[!htb]
\centering \caption{Description of the designed features as well as the feature sets for experiment}\label{feat}
\begin{tabular}{p{2.9 cm}p{0.5 cm}p{0.5 cm}p{0.5 cm}p{0.5 cm}p{0.5 cm}p{0.5 cm}}\\
\toprule[0.5mm]
{\it  Feature sets (FSs) } & 1 & 2 & 3  & 4  & 5 & 6\\
\midrule[0.5mm]
 Num. of UL packets  & 1 &  1&1&1&1&1\\ 
  Num. of DL packets  & 1 & 0& 0&1&1&1\\ 
 Size of UL packets& 1&0&0&0&0&0\\
  Size of DL packets& 1&0&0&0&0&0\\
UL\slash DL packets& 1& 1&0&1&0&0\\
Comm. protocol (TCP/UDP)& 0& 0&0&0&0&1\\
 \bottomrule[0.5mm]
\end{tabular}
\end{table}

\subsection{The Experiment Setup: Prediction and Classification} \label{sec:setup}
\begin{figure}
    \centering
    \begin{subfigure}[t]{.45\textwidth}
       \includegraphics[width=1\columnwidth]{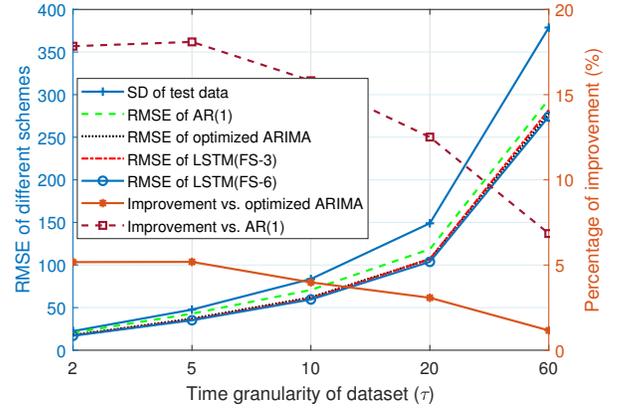}
                \caption{RMSE of prediction as a function of $\tau$ (time granularity of dataset)} 
                \label{rmselength}
     \end{subfigure}
     \begin{subfigure}[t]{.45\textwidth}
          \includegraphics[width=1\columnwidth]{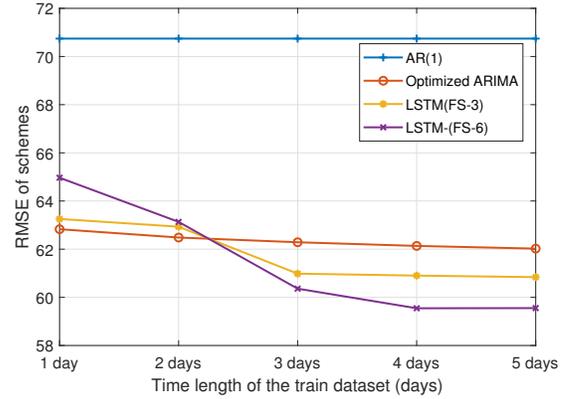}
                \caption{RMSE of prediction  of number of uplink packets as a function of length of the training dataset (as well as length of future prediction).} 
                \label{rmsetrain}
     \end{subfigure}
    \begin{subfigure}[t]{.45\textwidth}
        	\includegraphics[width=1\columnwidth]{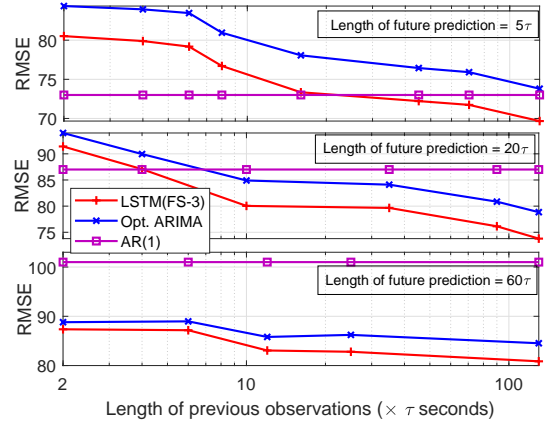}
	\caption{The RMSE performance of LSTM and ARIMA in short to long-range future traffic prediction ($\tau$=10 sec).} 
	\label{axkol}
     \end{subfigure}
     \caption{Performance analysis of traffic prediction.}
     \label{figfig}
\end{figure}

 The  experimental results in the following section are presented within two categories, i.e. i) prediction of number of packet arrivals in future  time intervals, and ii) classification of  applications which are generating the traffic, in order to answer the first two research questions raised in Section~\ref{sec:introduction}. In the first category, we did a comprehensive set of Monte Carlo MATLAB simulations \cite{mont},  over the dataset, for different lengths of the training sets, length of future prediction, feature sets used in learning and prediction, and etc. For example, each RMSE result in Fig. \ref{rmsetrain} for each scheme has been derived by averaging  over 37 experiments. In each experiment,  each scheme is trained using a train set starting from a random point of the dataset, and then is tested over 2000 future time intervals after the training set.   For the classification performance evaluation, we have leveraged 16 labeled datasets, each containing traffic from 4 mobile applications. Then, we construct 16 tests, in each test, one dataset is used for performance evaluation.  The notation of  schemes used in the experiments, extracted from the basic ARIMA and LSTM methods described in Section~\ref{sec:methods}, is as follows: (i) AR(1), which represents predicting the traffic  based on the last observation; (ii) optimized ARIMA, in which the number of lags and coefficients of ARIMA are optimized using a grid search for RMSE minimization; and (iii) LSTM(FS-$x$), in which FS-$x$ for $x\in\{1,\cdots,6\}$ represents the feature set used in the LSTM prediction/classification tool. 
 The overall configuration of experiments could be found in Table~\ref{exp}. 
  
 \begin{table}[!htb]
\centering \caption{Configuration of the experimental setup }\label{exp}
\begin{tabular}{p{4 cm} p{4 cm}}\\
\toprule[0.5mm]
{\it Parameters }& {\it Description}\\
\midrule[0.5mm]
Traffic type&  cellular traffic\\
Capture point&  IP layer, at the device side\\
Length of dataset& 48 days traffic\\
RNN  for prediction&  3-layer: [LSTM, fully connected, regression]\\
RNN  for classification&  3-layer: [LSTM, fully connected, softmax]\\
Granularity of prediction: $\tau$ &   default: 10 sec\\
Number of hidden elements&  100\\
 \bottomrule[0.5mm]
\end{tabular}
\end{table}

\smallskip
\noindent \textbf{Reproducibility of the results}
The experiments could be reproduced using the dataset available at the supporting Github repository \cite{GHAA}.

\subsection{Prediction and Classification Results   } \label{sec:results}

In this section, we present the prediction and classification performance results in two subsections, including prediction of traffic intensity in future time intervals and classification of traffic. In the first subsection, root mean square error (RMSE) is chosen as the performance indicator, while in the last subsection, accuracy and recall are the performance indicators. 

\subsubsection{Prediction of traffic intensity}
In the following figure,  we want to check if LSTM can further outperform the benchmark schemes by increasing time granularity of the dataset, decreasing length of future observation, and increasing length of the training set. First, let us  investigate the performance impact of $\tau$, i.e. the time granularity of dataset. Fig. \ref{rmselength} represents the absolute (left $y$-axis) and rational (right $y$-axis) RMSE results for the proposed and benchmark schemes as a function of time granularity of dataset ($\tau$, the $x$-axis). One must further consider the fact that $\tau$ not only represents how fine we have access to the history of the traffic, but also represents the length of future prediction. It is clear that the best results for the lowest $\tau$, e.g. when $\tau=2$, the LSTM(FS-6) outperforms the optimized ARIMA by 5\% and the AR(1) by 18\%. One further observes that by increasing the $\tau$, not only the RMSE increases, but also the merits of leveraging predictors decrease, e.g. for $\tau=60$, LSTM(FS-6) outperforms AR(1) by 7\%. Further, Fig. \ref{rmsetrain} investigates the performance impact of training set length on the prediction. One observes that the LSTM(FS-6) with poor training (1 day) even performs worse than optimized ARIMA. However, as the length of training dataset increases, the RMSE performance for the LSTM predictors, especially for LSTM(FS-6) with further features, decreases significantly.

Now, we investigate strengths of different predictors in medium to long-term traffic prediction. Fig. \ref{axkol} represents the RMSE results for 3 different lengths of future predictions, i.e. 50 sec (top), 200 sec (middle), and 600 sec (down). The $x$-axis represents the length of previous observations, i.e. it represents the number of observations just before the test window, which are available to be used by the trained model. The square-marked curve represents the results for AR(1), i.e. the case in which estimation is made based on the last observation. One observes that for medium-range future prediction, AR(1) outperforms the others when the number of previous observations is less than a threshold value, e.g. approximately 15 observations for $5\tau$-length future observations. Beyond this threshold value, we observe that LSTM outperforms the AR(1). Furthermore, we observe that this threshold value is dependent on the length of future prediction because in the middle and bottom figures, the LSTM predictor outperforms the others with the threshold value of 4 and 1 previous observations, respectively. The results further indicate that the optimized ARIMA, which has been optimized for traffic prediction in the next interval, loses its performance in longer ranges of future prediction, i.e. it is worse than AR(1) in some circumstances. \textcolor{black}{Moreover, it is observed when the length of future traffic prediction increases from $5\tau$ to $60\tau$ the RMSE is enhanced by just about $10\%$ which means there is no need for the high-frequent traffic prediction. So, the proposed model would be efficient enough to match the requirements for the scheduling of the DRX.} Finally, as observed in Fig. \ref{rmselength}, the relative performance of LSTM to AR(1) and ARIMA is highly dependent on the feature set used in training, and hence, the threshold value for LSTM decreases by incorporating further features.

\begin{figure}[!htb]
	\centering
	\includegraphics[width=1\columnwidth]{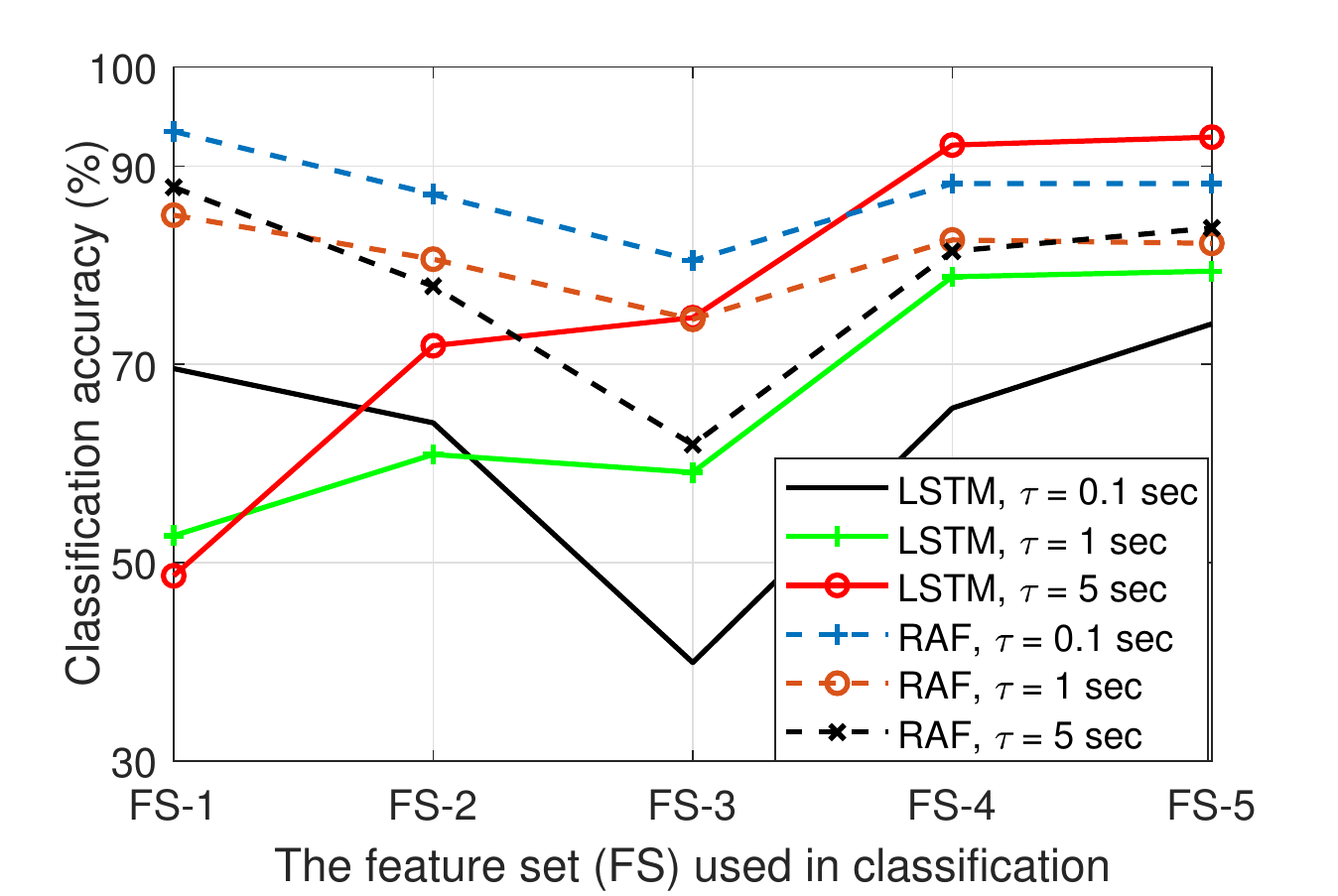}
	\caption{Accuracy of classification by LSTM and RAF as a function of the feature set used in the experiment.} 
	\label{accFSs}
\end{figure} 

\subsubsection{Classification of traffic}
 Finally, we investigate leveraging machine learning schemes for classification of the application generating the cellular traffic in this subsection. For the classification purpose, a controlled experiment at the  user-side has been carried out in which, 4 popular applications including surfing, video calling, voice calling, and video streaming have been used by the user. Fig. \ref{accFSs} represents the overall accuracy of classification for different feature sets used in the machine learning tool.  One observes that the LSTM(FS-5) and LSTM(FS-4) outperform the others significantly in the accuracy  of classification. Furthermore, in this figure,  3 curves for different lengths of the test data, to be classified, have been depicted. For example, when the length of the test data is 0.1 sec, the time granularity of dataset ($\tau$) is 0.1 sec, and  we also predict labels of intervals of length 0.1 sec.  For the LSTM-based solution, as the length of $\tau$ increases, the classification performance increases for feature sets which leverage less number of features. One reason could be that with a longer $\tau$, they have more evidence from the data in the test set to be matched to each class. This pattern is different in FS-1, which includes almost all features, and for the RAF solution. For the RAF, the reason could consist in the fact that is much simpler than the LSTM solution and cannot handle more variety and complexity in the dataset with a longer $\tau$. 
 
\begin{figure}[!htb]
 	\centering
 	\includegraphics[width=1\columnwidth]{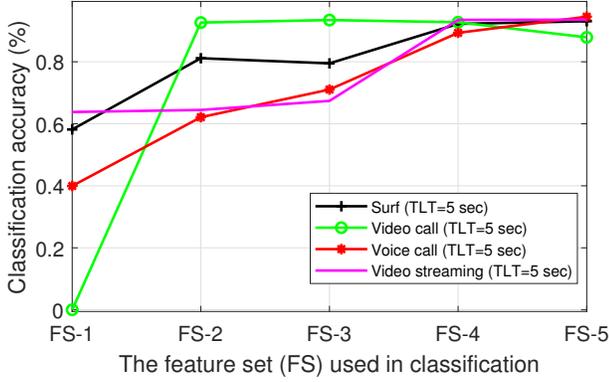}
 	\caption{Per application accuracy of classification as a function of the feature set used in the experiment.} 
 	\label{accperap}
 \end{figure}

 To further observe the recall of classification for different applications, Fig. \ref{accperap} represents the accuracy results per each application. One observes that the LSTM(FS-4) and LSTM(FS-5) outperform the others. It is also insightful that adding the ratio of uplink to downlink packets to FS-5, and hence constructing FS-4 (based on Table \ref{feat}), can make the prediction performance fairer for different applications. It is further insightful to observe that the choice of feature set to be used  is sensitive to the application used in the traffic dataset. In other words, FS-3, which benefits from one feature, outperforms the others in the accuracy of classification for video calling, while it results in classification error for other traffic types.

 \begin{figure*}[!htb]
 	\centering
 	\includegraphics[width=6in]{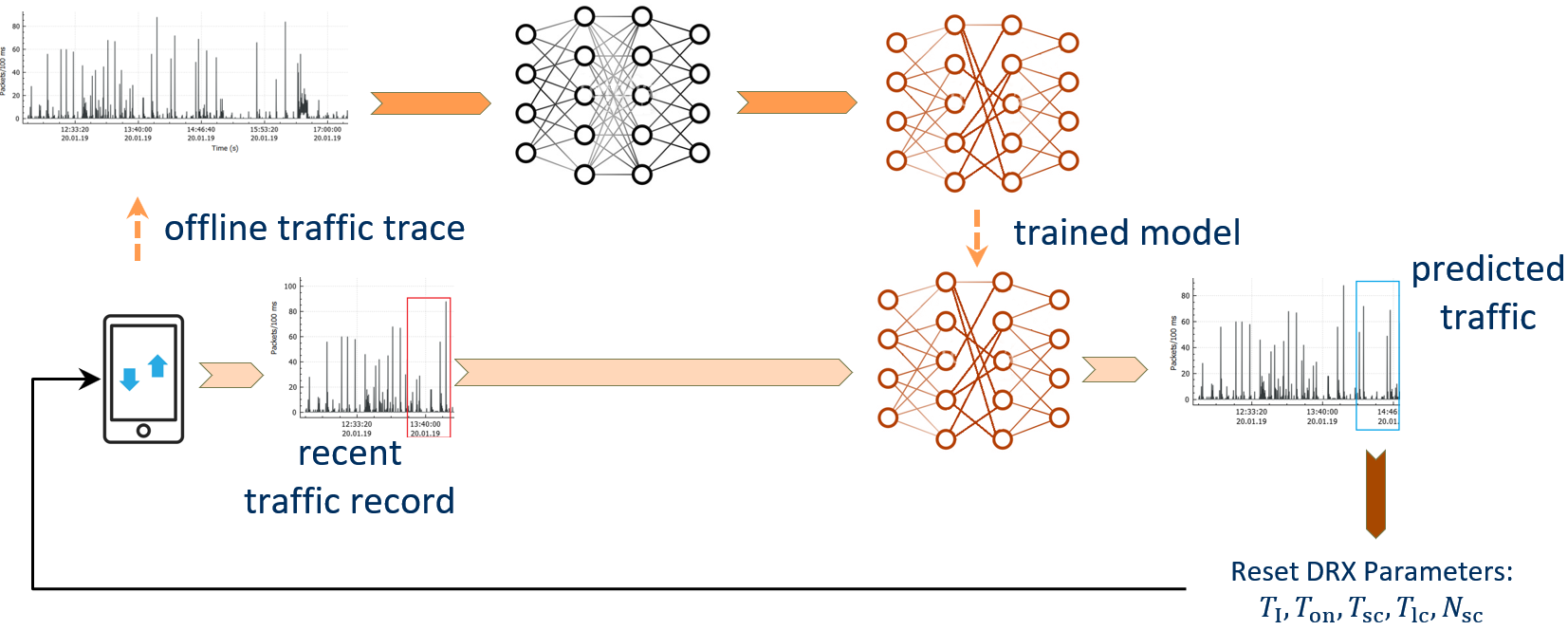}
 	\caption{Structure of the prediction-powered DRX parameter adaptation scheme.} 
 	\label{ovs}
 \end{figure*}

 \begin{algorithm}[htb]
 	 Initialize: DRX parameters and Prediction vector\; 
 	 Given: Trained traffic prediction model: $\mathcal F$\; 
 	 \For{$t=1,2,\cdots$}{
 		\If{need for change of DRX parameters}{
 			-Get recent traffic history: $\{\textbf X_1^{\textbf{s}_1}(t\text{-1}),\textbf X_2^{\textbf{s}_1}(t\text{-1}),\textbf X_3^{\textbf{s}_1}(t\text{-1}),\textbf X_4^{\textbf{s}_1}(t\text{-1}),$\\
 			$\sum\limits_{k=1}^{10} \textbf X_k^{\textbf{s}_1}(t\text{-1}), \sum\limits_{k=1}^{100}\textbf X_k^{\textbf{s}_1}(t\text{-1}), \sum\limits_{k=1}^{1000}\textbf X_k^{\textbf{s}_1}(t\text{-1})\}\triangleq \mathbb X$\;
 			-Derive traffic prediction:\\
 			$\mathcal F(\textbf X)\to \{\tilde{\textbf {X}}_\text{-1}^{\textbf{s}_2}(t),\tilde{\textbf {X}}_\text{-2}^{\textbf{s}_2}(t),\tilde{\textbf {X}}_\text{-3}^{\textbf{s}_2}(t),\tilde{\textbf {X}}_\text{-4}^{\textbf{s}_2}(t),$\\
 			$\sum\limits_{k=1}^{10} \tilde{\textbf {X}}_{\text{-}k}^{\textbf{s}_2}(t),\sum\limits_{k=1}^{100}\tilde{\textbf {X}}_{\text{-}k}^{\textbf{s}_2}(t), \sum\limits_{k=1}^{1000}\tilde{\textbf {X}}_{\text{-}k}^{\textbf{s}_2}(t)\}\triangleq\mathbb{\tilde{X}}$\;
 			$\sum\limits_{k=1:5} \mathbb {\tilde{X}}(k)\to \tilde{x}_{\text{short}}$\;
 			$\sum\limits_{k=6:7} \mathbb{ \tilde{X}}(k)\to \tilde{x}_{\text{long}}$\;
 			-Make decision for DRX-set:\\
 			{DRX-set~$=\mathcal H(\tilde{x}_{\rm{long}},\tilde{x}_{\rm{short}})$\;}
 	}}
 	\caption{The ML-based DRX  adaptation }\label{ho}
 \end{algorithm}

 \begin{table*}[!htb]
 	\centering \caption{Parameters for evaluation of traffic prediction-powered DRX parameter adaptation}\label{sim31}
 	\begin{tabular}{p{6 cm} p{8cm}}\\
 		\toprule[0.5mm]
 		{\it Parameters }&{\it Value}\\
 		\midrule[0.5mm]
 		Number of users& 10\\
 		Communications and radio resources& downlink service, 5 carriers, 1 Mbps data rate per carrier\\
 		Simulation time & 175 minutes (real traffic)\\
 		System BW&5 MHz\\ 
 		Power consumption in receiving data& 200 mW\\ 
 		Power consumption in active mode& 100 mW\\ 
 		Power consumption in sleep mode& 10 mW\\ 
 		DRX parameter set: & \\
 		$[T_\text{I} (\text{ms}),T_\text{on}(\text{ms}),T_\text{sc}(\text{ms}),N_\text{sc},T_\text{lc}(\text{ms})]$ & $\text{set 1}$=[02,01,05,10,015], $\text{set 2}$=[02,01,10,01,050],\\
 		& $\text{set 3}$=[10,03,04,20,010], $\text{set 4}$=[10,03,04,10,050]\\
 		Implemented scheme 1& most energy saving (DRX-set 2)\\
 		Implemented scheme 2& ML-based (probabilistic selection of DRX-set 1-4) \\
 		Implemented scheme 3& least delay (DRX-set 3)\\
 		\bottomrule[0.5mm]
 	\end{tabular}
 \end{table*}

 \section{ML-Powered DRX Adaptation}\label{drxa}
 In this section, we present our prediction-powered DRX adaptation algorithm and present its performance evaluation results. 
 
 \subsection{The DRX Parameter Sets}
 
 Each DRX set includes $[T_\text{I} (\text{ms}), T_\text{on}(\text{ms}), T_\text{sc}(\text{ms}), N_\text{sc}, T_\text{lc}(\text{ms})]$, where the first one is the inactivity timer, the second one is the on-time during a cycle, the third one is the length of short DRX cycle, the forth one is the number of short DRX cycles before a long DRX cycle, and the last one is the length of a long DRX cycle.  
 Our aim here is to dynamically attribute one of these DRX sets to the user based on its current and predicted traffic in future. In the following, we present an algorithm that predicts the future traffic, and based on this prediction, selects the best DRX set dynamically.

 \subsection{The DRX-adaptation Algorithm}
\begin{figure}[!htb]
 	\centering
 	\includegraphics[width=1\columnwidth]{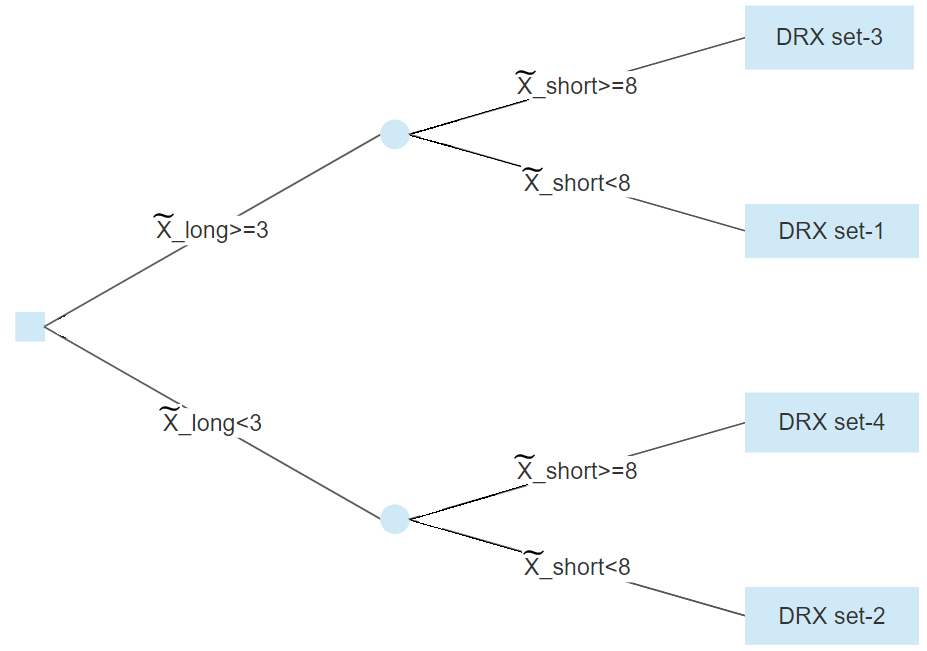}
 	\caption{The trained decision tree algorithm used as the mapping function $\mathcal H(\tilde{x}_{\rm{long}},\tilde{x}_{\rm{short}})$ in our simulations.} 
 	\label{ho1}
 \end{figure}

 The overall structure of the proposed solution for dynamic DRX parameter adaptation to user traffic has been depicted in Fig. \ref{ovs}. In this figure, one sees that the trace of traffic is used first for offline training of a ML network. Then the trained model is used online for traffic prediction based on the recent traffic arrival records. Then, the traffic prediction results is used in adaptation of $T_\text{I}$, $T_\text{on}$, $T_\text{sc}$, $N_\text{sc}$, and $T_\text{lc}$ parameters.
 One must note that before training and prediction, the traffic arrival data (size of traffic arrival) is quantized based on the following vector, [0,50,100,500,1000,5000,10000,50000,100000] bytes,  and is labelled by 1 to 9 based on its quantized value. For example, if the size of traffic arrival in the next TTI is 10 bytes, it is labelled 1. 
 Our algorithm for dynamic DRX parameter allocation has been reflected in Algorithm \ref{ho}. In this algorithm, $t$ denotes the transmission time interval (TTI), which is 1 msec in the existing networks. Because the DRX parameters usually change in the order of seconds and minutes, there is no need to go through the algorithm in all TTIs, and hence, we have the IF condition at the beginning of the algorithm. Furthermore, the recent traffic record, the line after the first IF structure, contains traffic arrival in last 4 TTIs, and sum of arrivals in last 10 TTIs, 100 TTIs, and 1000 TTIs\footnote{One must note that the above mentioned summations are summation of labels of traffic arrivals, as we discussed before that size of traffic arrival is quantized and labelled from 1 to 9, based on the quantization vector.}. In the next step of the algorithm, we derive predictions of the arrival in the next 4 TTIs, as well as the sum of arrivals in the next 10, 100, and 1000 TTIs. Finally, we sum the predicted traffic arrivals in two groups, namely short and long future. $\tilde{x}_{\text{short}}$ includes sum of predictions for the first 4 TTIs as well as the sum of 10 future TTIs, while $\tilde{x}_{\text{long}}$ includes sum of future arrivals in the 100 and 1000 TTIs. Now, we attribute the DRX parameter set to the device based on the values of  $\tilde{x}_{\text{long}}$ and  $\tilde{x}_{\text{short}}$, as well as the predicted application behind the traffic, using a mapping function, called $\mathcal H$. One exemplary mapping function, which is used in our simulations for web surfing, has been depicted in Fig. \ref{ho1} as a function of $\tilde{x}_{\rm{long}}$ and $\tilde{x}_{\rm{short}}$, i.e. denoted by $\mathcal H(\tilde{x}_{\rm{long}},\tilde{x}_{\rm{short}})$. This figure represents a decision tree, trained over out application traffic, which maps the results the prediction of traffic to the DRX sets.  It is possible to embed this step also in the traffic prediction module, i.e. the neural network, but here we have extracted this  step from the overall ML-powered DRX adaptation solution for the transparency of the solution, i.e. to see how machine reasoning carries out the DRX adaptation.  
 
 % \begin{algorithm}[htb]
 %	 Given: Prediction features $\tilde{x}_{\rm{long}}$; $\tilde{x}_{\rm{short}}$\; 
% 	\uIf{$\tilde{x}_{\rm{short}}<8 ~\&~ \tilde{x}_{\rm{long}}\ge3 $}{DRX-set~$=1$\;}
 %			\uElseIf{$\tilde{x}_{\rm{short}}<8 ~\&~ \tilde{x}_{\rm{long}}<3 $}{DRX-set~$=2$\;}
 %			\uElseIf{$\tilde{x}_{\rm{short}}\ge 8 ~\&~ \tilde{x}_{\rm{long}}\ge3 $}{DRX-set~$=3$\;}  
 %			\ElseIf{$\tilde{x}_{\rm{short}}\ge 8 ~\&~ \tilde{x}_{\rm{long}}<3 $}{DRX-set~$=4$\;} 
 %	\caption{Pseudo-code for the exemplary mapping function $\mathcal H(\tilde{x}_{\rm{long}},\tilde{x}_{\rm{short}})$}\label{ho1}
 %\end{algorithm}
 
 \subsection{The Simulation Setup }
 The simulation setup has been described in Table \ref{sim31}. Here, we have considered a network of 10 users, and 5 downlink carriers. The capacity of each carrier is 1 Mbps. Activity of user devices is based on 10 real dataset of 175 minutes long, and 10 neural networks are trained to capture the characteristics of traffic of each user. For simulations, we  consider 4 sets of DRX parameters to be used by the device in this work, as follows:  
 $\text{DRX-set 1}$=[02,01,05,10,015], $\text{DRX-set 2}$=[02,01,10,01,050],
 $\text{DRX-set 3}$=[10,03,04,20,010], 
 $\text{DRX-set 4}$=[10,03,04,10,050]. 
 It is clear that one can extend this set and define further DRX sets covering different values for each of timers to reach more delay-energy tradeoffs. 
 
 \subsection{The Results }
 
\begin{figure}[!htb]
 	\centering
 	\includegraphics[width=1\columnwidth]{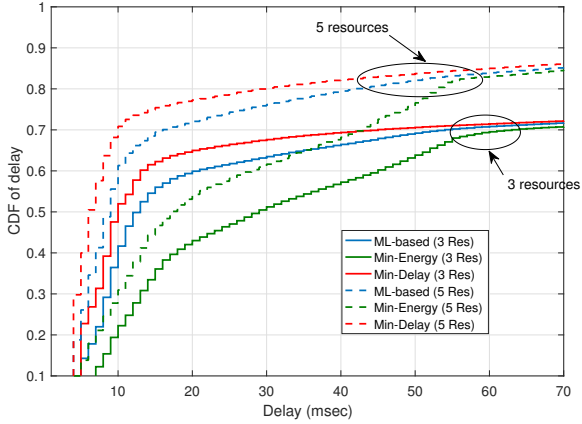}
 	\caption{The CDF of experienced delay by data packets during the DRX.} 
 	\label{dels}
 \end{figure}

\begin{figure}[!htb]
 	\centering
 	\includegraphics[width=1\columnwidth]{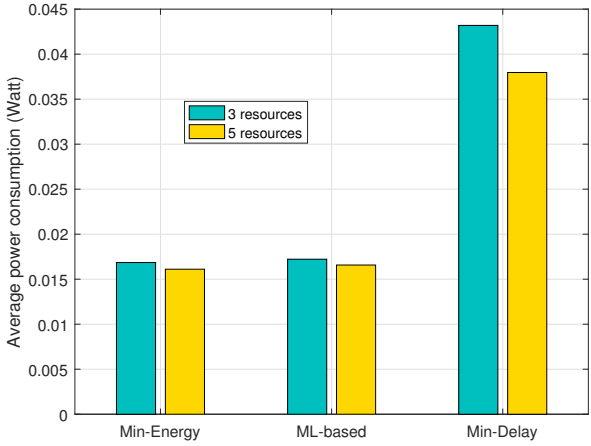}
 	\caption{Average power consumption in three hours by the three schemes during the DRX.} 
 	\label{en}
 \end{figure}

 Fig. \ref{dels} represents the CDF delay results for the case 3 and  5 carriers are used in serving downlink traffic of 10 users in the service area. The three schemes here are min delay, min energy, and the ML-based prediction-powered solution. One observes that the prediction-powered solutions offer a tradeoff between delay and energy, which can be tuned by changing the decision boundaries in Fig. \ref{ho1}, i.e. 3 and 8. Fig. \ref{en} represents the average power consumption results. One observes that device power consumption in the Min-Delay schemes is more than twice the power consumption in the most energy-saving mode (which performs bad in delay). Interestingly, one observes that the proposed scheme achieves a very good performance in energy consumption, very close to the optimal value, while its delay performance in Fig. \ref{dels} is also acceptable.

One can see that from energy efficiency in Fig. \ref{en}, ML-based solution and Min-Energy are almost the same. However, from the Fig. \ref{dels} we observe that there are operation regions where ML-based solution  with 3 resources works better  than Min-Energy solution with 5 resources. Then, our solution brings spectrum efficiency as well.

\subsection{Discussion  on the Results } \label{sec:discussion}
In this subsection, we  discuss how the raised research questions throughout this work have been answered in our analyses and experiments. 

\noindent
{\textbf{First}}, how accurately can we estimate the intensity of traffic in the next time intervals? The experimental results represent that the accuracy of prediction strongly depends on the length of training dataset, time granularity of dataset, length of future prediction, mode of activity of the user (standard deviation of test dataset), and the feature set used in the learning scheme. The results, for example, indicate that the proposed LSTM(FS-3) is performing approximately 5\% better than the optimized ARIMA, and 18\% better than AR(1) for $\tau=10$ sec.  The results further indicated that the performance of LSTM could be further improved by designing more features related to the traffic, e.g. the protocol in use for packets, and the ratio of uplink to downlink packets.

\noindent
{\textbf{Second}}, how accurately can we estimate the application which is generating the traffic? The experimental results represented the facts that, first, accuracy and recall performance of classification is highly dependent on the feature set used in the classification. For example, a feature set that can achieve an accuracy of 90\% for classification of one application may result in a recall of 10\% for another application. Then, the choice of feature set should be in accordance with the set of applications used by the user.  
Second, if we can tolerate delay in the decision, e.g. 5 sec, the classification performance will be much more accurate when we gather more information and decide on longer time intervals. The overall accuracy performance for different applications using the developed classification tool is approximately 90\%.

\noindent
{\textbf{Third}}, how can one employ the learnt model for data prediction in DRX optimization?  In the proposed solution, using a long-enough trace of traffic, a ML-powered traffic arrival prediction is trained, and then, at each decision epoch, using a short history of traffic in the previous time instances and the learnt-model, DRX parameters are adapted to the traffic prediction. The simulation results, leveraging real traffic arrival traces, confirmed that adaptation of DRX parameters by online prediction of future traffic  provides much more energy saving at low latency cost in comparison with the legacy cell-wide DRX parameter adaptation. 

\textcolor{black}{\subsection{Future Work} The research methodology in this work is evaluation of the AI algorithms and their performance in traffic prediction. Moreover, as a usecase, we show that this prediction could be useful in configuration of device's parameters, e.g., DRX setups. Thus, the analysis for DRX is not including further investigation of impact of accuracy, recall, F1-score on the energy saving, as well as, on reachability of the device. Due to the page limit, we restrict the scope of DRX analysis to a usecase study in this manuscript, and further analysis is left for the future publication.}

%%%%%%%%%%%%%%%%%%%%%%%%%%%%%%%%%%%%%%%%%%%%%%%%
%%%%%%%%%%%%%%%%%%%%%%%%%%%%%%%%%%%%%%%%%%%%%%%%

\section{Concluding Remarks} \label{sec:conclusions}
In this work, the feasibility of per-user traffic prediction for cellular networks has been  investigated. Towards this end, a framework for cellular traffic prediction has been introduced, which leverages statistical/machine learning units for traffic classification and prediction. A comprehensive comparative analysis of prediction tools based on statistical learning, ARIMA, and the one based on machine learning, LSTM, has been carried out, under different traffic circumstances and design parameter selections.  The LSTM model, in particular when augmented by additional features like the ratio of uplink to downlink packets and the communication protocol used in prior packet transfers, exhibited demonstrable improvement over the ARIMA model for  future traffic predictions. Furthermore,  usefulness of the developed LSTM model for classification of cellular traffic has been investigated, where the results represent high sensitivity of accuracy and recall of classification to the feature set in use. Finally, the learnt models for traffic prediction has been employed in online adaptation of DRX communications' parameters of cellular users. The results show that adaptation of DRX parameters by online prediction of future traffic  provides much more energy saving at low latency cost in comparison with the legacy cell-wide DRX parameter adaptation. 

\textcolor{black}{Although there are much more advanced algorithms in the literature, such as multitask-learning or transfer-learning, able to extract further patterns from the traffic, in this manuscript, our main purpose was on doing a sanity check to see if AI can help in configuring communication parameters of devices for saving, and if so, what are the main design principles and bounds of operation. Hence, we have limited the scope to a simple AI algorithm, available on most platforms, and easy to reproduce, to do this check and investigation. While more sophisticated algorithms are powerful in specific tasks once trained, they are less suitable to adapt to changes in the dataset and applicable to a wide set of services. On the other hand, while a simple trained ARIMA scheme is less capable of deriving deep patterns, it performs quite well over a wide set of datasets. At the same time, a scheme based on neural networks performs pretty well on a dataset that has been trained, but works poorly in a new dataset.} The presented results in this work promote further research on user traffic prediction and  classification both for more efficient radio resource provisioning at the network side and further energy saving at the user side through optimized DRX parameter adaptation.

\ifCLASSOPTIONcaptionsoff
  \newpage
\fi

\bibliographystyle{IEEEtran}
\bibliography{paperbib}

\begin{thebibliography}{10}
\providecommand{\url}[1]{#1}
\csname url@samestyle\endcsname
\providecommand{\newblock}{\relax}
\providecommand{\bibinfo}[2]{#2}
\providecommand{\BIBentrySTDinterwordspacing}{\spaceskip=0pt\relax}
\providecommand{\BIBentryALTinterwordstretchfactor}{4}
\providecommand{\BIBentryALTinterwordspacing}{\spaceskip=\fontdimen2\font plus
\BIBentryALTinterwordstretchfactor\fontdimen3\font minus
  \fontdimen4\font\relax}
\providecommand{\BIBforeignlanguage}[2]{{%
\expandafter\ifx\csname l@#1\endcsname\relax
\typeout{** WARNING: IEEEtran.bst: No hyphenation pattern has been}%
\typeout{** loaded for the language `#1'. Using the pattern for}%
\typeout{** the default language instead.}%
\else
\language=\csname l@#1\endcsname
\fi
#2}}
\providecommand{\BIBdecl}{\relax}
\BIBdecl

\bibitem{Sad6g}
W.~{Saad}, M.~{Bennis}, and M.~{Chen}, ``A vision of {6G} wireless systems:
  Applications, trends, technologies, and open research problems,'' \emph{IEEE
  Network}, vol.~34, no.~3, pp. 134--142, 2020.

\bibitem{RaUrllc}
A.~Azari, M.~Ozger, and C.~Cavdar, ``Risk-aware resource allocation for
  {URLLC}: Challenges and strategies with machine learning,'' \emph{IEEE
  Communications Magazine}, 2019.

\bibitem{Db6g}
M.~{Chen}, U.~{Challita}, W.~{Saad}, C.~{Yin}, and M.~{Debbah}, ``Artificial
  neural networks-based machine learning for wireless networks: A tutorial,''
  \emph{IEEE Communications Surveys Tutorials}, vol.~21, no.~4, pp. 3039--3071,
  2019.

\bibitem{RnnScale}
I.~Alawe \emph{et~al.}, ``Smart scaling of the {5G} core network: an
  {RNN}-based approach,'' in \emph{Globecom 2018-IEEE Global Communications
  Conference}, 2018, pp. 1--6.

\bibitem{DrlBsSleep}
J.~Ye and Y.-J.~A. Zhang, ``{DRAG}: Deep reinforcement learning based base
  station activation in heterogeneous networks,'' \emph{IEEE Transactions on
  Mobile Computing}, vol.~19, no.~9, pp. 2076--2087, 2020.

\bibitem{UavML}
Q.~{Zhang}, M.~{Mozaffari}, W.~{Saad}, M.~{Bennis}, and M.~{Debbah}, ``Machine
  learning for predictive on-demand deployment of uavs for wireless
  communications,'' in \emph{2018 IEEE Global Communications Conference
  (GLOBECOM)}, 2018, pp. 1--6.

\bibitem{SelfMl}
A.~Azari and C.~Cavdar, ``Self-organized low-power {IoT} networks: A
  distributed learning approach,'' \emph{IEEE Globecom 2018}, 2018.

\bibitem{moradi2021improving}
F.~Moradi, ``Improving {DRX} performance for emerging use cases in {5G},''
  Ph.D. dissertation, Lund University, 2021.

\bibitem{ze}
T.~Haque, H.~Elkotby, P.~Cabrol, R.~Pragada, and D.~Castor, ``A supplemental
  zero-energy downlink air-interface enabling 40-year battery life in {IoT}
  devices,'' in \emph{GLOBECOM 2020-2020 IEEE Global Communications
  Conference}.\hskip 1em plus 0.5em minus 0.4em\relax IEEE, 2020, pp. 1--6.

\bibitem{TiSerNn}
A.~Tealab, ``Time series forecasting using artificial neural networks
  methodologies: A systematic review,'' \emph{Future Computing and Informatics
  Journal}, vol.~3, no.~2, pp. 334 -- 340, 2018.

\bibitem{Hyb}
G.~P. Zhang, ``Time series forecasting using a hybrid {ARIMA} and neural
  network model,'' \emph{Neurocomputing}, vol.~50, pp. 159--175, 2003.

\bibitem{CellTrChar}
Y.~Zhang and A.~{\AA}rvidsson, ``Understanding the characteristics of cellular
  data traffic,'' in \emph{Proceedings of the 2012 ACM SIGCOMM workshop on
  Cellular networks: operations, challenges, and future design}, 2012, pp.
  13--18.

\bibitem{azari2019user}
A.~Azari, P.~Papapetrou, S.~Denic, and G.~Peters, ``User traffic prediction for
  proactive resource management: Learning-powered approaches,'' \emph{IEEE
  Globecom 2019}, 2019.

\bibitem{azari2019cellular}
------, ``Cellular traffic prediction and classification: A comparative
  evaluation of {LSTM and ARIMA},'' \emph{Lecture Notes in Computer Science,
  vol 11828. Springer}, 2019.

\bibitem{6GArch_automation}
A.~H. Sodhro, S.~Pirbhulal, Z.~Luo, K.~Muhammad, and N.~Z. Zahid, ``Toward 6g
  architecture for energy-efficient communication in iot-enabled smart
  automation systems,'' \emph{IEEE Internet of Things Journal}, vol.~8, no.~7,
  pp. 5141--5148, 2021.

\bibitem{netAutomation}
D.~Rafique and L.~Velasco, ``Machine learning for network automation: Overview,
  architecture, and applications [invited tutorial],'' \emph{J. Opt. Commun.
  Netw.}, vol.~10, no.~10, pp. D126--D143, Oct 2018.

\bibitem{bookrima}
P.~J. Brockwell, R.~A. Davis, and M.~V. Calder, \emph{Introduction to time
  series and forecasting}.\hskip 1em plus 0.5em minus 0.4em\relax Springer,
  2002, vol.~2.

\bibitem{Seq2Seq}
J.~Rebane, I.~Karlsson, S.~Denic, and P.~Papapetrou, ``Seq2seq {RNN}s and
  {ARIMA} models for cryptocurrency prediction: A comparative study,'' 2018.

\bibitem{LstmTrafficRaw}
H.~D. Trinh, L.~Giupponi, and P.~Dini, ``Mobile traffic prediction from raw
  data using {LSTM} networks,'' in \emph{2018 IEEE 29th Annual International
  Symposium on Personal, Indoor and Mobile Radio Communications (PIMRC)}, 2018,
  pp. 1827--1832.

\bibitem{SpTmBigDL}
J.~Wang \emph{et~al.}, ``Spatio-temporal modeling and prediction in cellular
  networks: A big data enabled deep learning approach,'' in \emph{IEEE INFOCOM
  2017-IEEE Conference on Computer Communications}, 2017, pp. 1--9.

\bibitem{HybArimaLstm}
H.~K. Choi, ``Stock price correlation coefficient prediction with
  {ARIMA}-{LSTM} hybrid model,'' \emph{arXiv preprint arXiv:1808.01560}, 2018.

\bibitem{ArimaLstm}
T.~Skehin, M.~Crane, and M.~Bezbradica, ``Day ahead forecasting of {FAANG}
  stocks using {ARIMA}, {LSTM} networks and wavelets,'' in \emph{CEUR Workshop
  Proceedings}, 2018.

\bibitem{PersonDemand}
T.~Chen, B.~Keng, and J.~Moreno, ``Multivariate arrival times with recurrent
  neural networks for personalized demand forecasting,'' \emph{arXiv preprint
  arXiv:1812.11444}, 2018.

\bibitem{NeuTM}
A.~Azzouni and G.~Pujolle, ``Neutm: A neural network-based framework for
  traffic matrix prediction in {SDN},'' in \emph{NOMS 2018-2018 IEEE/IFIP
  Network Operations and Management Symposium}, 2018, pp. 1--5.

\bibitem{DlForecast}
C.-W. Huang, C.-T. Chiang, and Q.~Li, ``A study of deep learning networks on
  mobile traffic forecasting,'' in \emph{2017 IEEE 28th Annual International
  Symposium on Personal, Indoor, and Mobile Radio Communications (PIMRC)},
  2017, pp. 1--6.

\bibitem{EntropyCellular}
R.~Li \emph{et~al.}, ``The prediction analysis of cellular radio access network
  traffic: From entropy theory to networking practice,'' \emph{IEEE
  Communications Magazine}, vol.~52, no.~6, pp. 234--240, 2014.

\bibitem{NetDemand}
H.~Assem, B.~Caglayan, T.~S. Buda, and D.~O’Sullivan, ``St-dennetfus: A new
  deep learning approach for network demand prediction,'' in \emph{Joint
  European Conference on Machine Learning and Knowledge Discovery in
  Databases}, 2018, pp. 222--237.

\bibitem{SpTmTraffic}
C.~Qiu, Y.~Zhang, Z.~Feng, P.~Zhang, and S.~Cui, ``Spatio-temporal wireless
  traffic prediction with recurrent neural network,'' \emph{IEEE Wireless
  Communications Letters}, vol.~7, no.~4, pp. 554--557, 2018.

\bibitem{ModelLstmDnn}
G.~Lai, W.-C. Chang, Y.~Yang, and H.~Liu, ``Modeling long-and short-term
  temporal patterns with deep neural networks,'' in \emph{The 41st
  International {ACM SIGIR} Conference on Research \& Development in
  Information Retrieval}, 2018, pp. 95--104.

\bibitem{deepcl}
S.~Rezaei and X.~Liu, ``Deep learning for encrypted traffic classification: An
  overview,'' \emph{arXiv preprint arXiv:1810.07906}, 2018.

\bibitem{cls2006}
N.~Williams, S.~Zander, and G.~Armitage, ``A preliminary performance comparison
  of five machine learning algorithms for practical {IP} traffic flow
  classification,'' \emph{ACM SIGCOMM Computer Communication Review}, vol.~36,
  no.~5, pp. 5--16, 2006.

\bibitem{lopez}
M.~Lopez-Martin, B.~Carro, A.~Sanchez-Esguevillas, and J.~Lloret, ``Network
  traffic classifier with convolutional and recurrent neural networks for
  internet of things,'' \emph{IEEE Access}, vol.~5, pp. 18\,042--18\,050, 2017.

\bibitem{RnnClass}
V.~Tong, H.~A. Tran, S.~Souihi, and A.~Melouk, ``A novel {QUIC} traffic
  classifier based on convolutional neural networks,'' in \emph{IEEE
  International Conference on Global Communications (GlobeCom)}, 2018, pp.
  1--6.

\bibitem{p1}
H.~{Ferng} and T.~{Wang}, ``Exploring flexibility of {DRX in LTE/LTE-A: Design
  of Dynamic and Adjustable DRX},'' \emph{IEEE Transactions on Mobile
  Computing}, vol.~17, no.~1, pp. 99--112, Jan 2018.

\bibitem{p2}
K.~{Wang}, X.~{Li}, H.~{Ji}, and X.~{Du}, ``Modeling and optimizing the {LTE}
  discontinuous reception mechanism under self-similar traffic,'' \emph{IEEE
  Transactions on Vehicular Technology}, vol.~65, no.~7, pp. 5595--5610, July
  2016.

\bibitem{p3}
A.~T. {Koc}, S.~C. {Jha}, R.~{Vannithamby}, and M.~{Torlak}, ``Device power
  saving and latency optimization in {LTE-A} networks through {DRX}
  configuration,'' \emph{IEEE Transactions on Wireless Communications},
  vol.~13, no.~5, pp. 2614--2625, May 2014.

\bibitem{p4}
S.~C. {Jha}, A.~T. {Koç}, and R.~{Vannithamby}, ``Optimization of
  discontinuous reception (drx) for mobile internet applications over lte,'' in
  \emph{2012 IEEE Vehicular Technology Conference (VTC Fall)}, Sep. 2012, pp.
  1--5.

\bibitem{p5}
J.~{Zhou}, G.~{Feng}, T.~P. {Yum}, M.~{Yan}, and S.~{Qin}, ``Online
  learning-based discontinuous reception (drx) for machine-type
  communications,'' \emph{IEEE Internet of Things Journal}, vol.~6, no.~3, pp.
  5550--5561, June 2019.

\bibitem{p6}
D.~{Corcoran}, L.~{Andimeh}, A.~{Ermedahl}, P.~{Kreuger}, and C.~{Schulte},
  ``Data driven selection of {DRX} for energy efficient {5G RAN},'' in
  \emph{2017 13th International Conference on Network and Service Management
  (CNSM)}, Nov 2017, pp. 1--9.

\bibitem{p7}
M.~L. Memon, M.~K. Maheshwari, D.~R. Shin, A.~Roy, and N.~Saxena, ``{Deep-DRX:
  A framework for deep learning--based discontinuous reception in 5G wireless
  networks},'' \emph{Transactions on Emerging Telecommunications Technologies},
  vol.~30, no.~3, p. e3579, 2019.

\bibitem{wu2021adaptive}
J.~Wu, B.~Yang, L.~Wang, and J.~Park, ``Adaptive {DRX} method for {MTC} device
  energy saving by using a machine learning algorithm in an {MEC} framework,''
  \emph{IEEE Access}, vol.~9, pp. 10\,548--10\,560, 2021.

\bibitem{ruiz2021drx}
D.~E. Ru{\'\i}z-Guirola, C.~A. Rodr{\'\i}guez-L{\'o}pez,
  S.~Montejo-S{\'a}nchez, R.~D. Souza, and M.~A. Imran, ``{DRX}-based
  energy-efficient supervised machine learning algorithm for mobile
  communication networks,'' \emph{IET Communications}, vol.~15, no.~7, pp.
  1000--1013, 2021.

\bibitem{sundararaju2020novel}
S.~C. Sundararaju, M.~Balasubramaniam, and D.~Das, ``Novel {C-DRX} mechanism
  for multi sim multi standby {UE}s in {5G and B5G} networks,'' in \emph{IEEE
  3rd 5G World Forum {(5GWF)}}, 2020, pp. 318--323.

\bibitem{box1976time}
G.~E. Box and G.~M. Jenkins, \emph{Time series analysis: forecasting and
  control, revised ed}.\hskip 1em plus 0.5em minus 0.4em\relax Holden-Day,
  1976.

\bibitem{mills1991time}
T.~C. Mills, \emph{Time series techniques for economists}.\hskip 1em plus 0.5em
  minus 0.4em\relax Cambridge University Press, 1991.

\bibitem{RAF_ref}
V.~K. Ayyadevara, ``Random forest,'' in \emph{Pro Machine Learning
  Algorithms}.\hskip 1em plus 0.5em minus 0.4em\relax Springer, 2018, pp.
  105--116.

\bibitem{DBLP:journals/corr/ChoMGBSB14}
\BIBentryALTinterwordspacing
K.~Cho, B.~van Merrienboer, {\c{C}}.~G{\"{u}}l{\c{c}}ehre, F.~Bougares,
  H.~Schwenk, and Y.~Bengio, ``Learning phrase representations using {RNN}
  encoder-decoder for statistical machine translation,'' \emph{CoRR}, vol.
  abs/1406.1078, 2014. [Online]. Available:
  \url{http://arxiv.org/abs/1406.1078}
\BIBentrySTDinterwordspacing

\bibitem{GHAA}
A.~Azari, ``Cellular traffic analysis,''
  \url{https://github.com/AminAzari/cellular-traffic-analysis}, {GitHub}
  repository, Accessed: 6/8/2021.

\bibitem{ng}
``Netguard - no-root firewall,''
  \url{https://play.google.com/store/apps/details?\\id=eu.faircode.netguard},
  {A}ccessed: 2021-06-08.

\bibitem{mont}
R.~Y. Rubinstein and D.~P. Kroese, \emph{Simulation and the Monte Carlo
  method}.\hskip 1em plus 0.5em minus 0.4em\relax John Wiley \& Sons, 2016,
  vol.~10.

\end{thebibliography}

\end{document}